\documentclass[preprint, aps, pre,longbibliography,showpacs]{revtex4-1} 
\usepackage{graphicx, bm, bbm,amsmath, amssymb, epsfig}

\def\a{\alpha}

\def\bX{{\bf X}}
\def\bF {{\bf f}}
\def\th {\theta}
\def\k {\kappa}
\def\hs{{\hat{\bf s}}}
\def\hn{{\hat{\bf n}}}
\def\ptX{\partial_t\bX}
\def\hptX{\widehat{\partial_t\bX}}
\def\ptx0{\partial_t x_0}
\def\pty0{\partial_t y_0}
\def\ptth0{\partial_t \theta_0}
\def\gpavg{\langle g'(s+U_wt)^2\rangle}
\def\gpavgs{\langle g'^2\rangle}

\newcommand{\bq}{\begin{equation}}
\newcommand{\eq}{\end{equation}}
\newcommand{\bqs}{\begin{equation*}}
\newcommand{\eqs}{\end{equation*}}
\newcommand{\bqa}{\begin{eqnarray}}
\newcommand{\eqa}{\end{eqnarray}}
\newcommand{\bqas}{\begin{eqnarray*}}
\newcommand{\eqas}{\end{eqnarray*}}

\def\Xint#1{\mathchoice
{\XXint\displaystyle\textstyle{#1}}%
{\XXint\textstyle\scriptstyle{#1}}%
{\XXint\scriptstyle\scriptscriptstyle{#1}}%
{\XXint\scriptscriptstyle\scriptscriptstyle{#1}}%
\!\int}
\def\XXint#1#2#3{{\setbox0=\hbox{$#1{#2#3}{\int}$ }
\vcenter{\hbox{$#2#3$ }}\kern-.6\wd0}}

\def\dashint{\Xint-}
\def\sg{\langle sg'\rangle}
\def\gp{\langle g'\rangle}
\def\mt{\mu_t}
\def\mf{\mu_f}
\def\mtmf{\frac{\mu_t}{\mu_f}}
\newcommand{\sgn}{\operatorname{sgn}}

\def\etal{{\em et al.\ }}

\newcommand{\Dfig}[2]{\includegraphics*[width=#2in]{#1.eps}}
\newenvironment{Eonefigs}[3]
{
        \begin{figure} [ht]
          \begin{center}
          \begin{tabular}{c}
              \Dfig{#1}{#3} \\
           \vspace{-.25in}
          \end{tabular}
         \caption{{\footnotesize #2}}
          \label{#1}
          \end{center}
        \vspace{-0in}
       \end{figure}
}{}

\newenvironment{Ethreefigs}[7]
{
        \begin{figure} [htpb]
          \begin{center}
          \setlength{\tabcolsep}{.1in}
          \begin{tabular}{ccc}
          \multicolumn{1}{l}{$\bf{(a)}$}&\multicolumn{1}{l}{$\bf{(b)}$}&\multicolumn{1}{l}{$\bf{(c)}$}\\
              \Dfig{#1}{#5} & \Dfig{#2}{#6}&\Dfig{#3}{#7} \\
                    \vspace{-.25in}
          \end{tabular}
          \caption{\footnotesize #4}
          \label{#1}\label{#2}\label{#3}
          \end{center}
        \vspace{-.25in}
        \end{figure}
}{}

\newenvironment{Efourfigs}[7]
{
        \begin{figure} [htpb]
          \begin{center}
           \setlength{\tabcolsep}{.1in}
          \begin{tabular}{cc}
          \multicolumn{1}{l}{$\bf{(a)}$}&\multicolumn{1}{l}{$\bf{(b)}$}\\
              \Dfig{#1}{#6} & \Dfig{#2}{#6} \\
              \multicolumn{1}{l}{$\bf{(c)}$}&\multicolumn{1}{l}{$\bf{(d)}$}\\
                     \Dfig{#3}{#7} & \Dfig{#4}{#7} \\
                   \vspace{-.25in}
          \end{tabular}
          \caption{{\footnotesize #5}}
          \label{#1}\label{#2}\label{#3}\label{#4}
          \end{center}
        \vspace{-.25in}
        \end{figure}
}{}
\newenvironment{ESixfigs}[8]
{
        \begin{figure} [htpb]
          \begin{center}
          \setlength{\tabcolsep}{.1in}
          \begin{tabular}{ccc}
          \multicolumn{1}{l}{$\bf{(a)}$}&\multicolumn{1}{l}{$\bf{(b)}$}&\multicolumn{1}{l}{$\bf{(c)}$}\\
              \Dfig{#1}{#8} & \Dfig{#2}{#8}&\Dfig{#3}{#8} \\
            \multicolumn{1}{l}{$\bf{(d)}$}&\multicolumn{1}{l}{$\bf{(e)}$}&\multicolumn{1}{l}{$\bf{(f)}$}\\
              \Dfig{#4}{#8} & \Dfig{#5}{#8}&\Dfig{#6}{#8} \\
                    \vspace{-.25in}
          \end{tabular}
          \caption{{\footnotesize #7}}
          \label{#1}\label{#2}\label{#3}\label{#4}\label{#5}\label{#6}
          \end{center}
        \vspace{-.25in}
        \end{figure}
}{}

\begin{document}
\title{Optimizing snake locomotion on an inclined plane}
\author{Xiaolin Wang$^{1,2,*}$, Matthew T. Osborne$^{1,3}$, and Silas Alben$^{1,*}$}
\affiliation{$^1$University of Michigan, $^2$Georgia Institute of Technology,
$^3$University of Toledo}
\email{xiaolinw@gatech.edu, alben@umich.edu}

\date{\today}

\begin{abstract}
We develop a model to study the locomotion of snakes on an inclined plane. We determine numerically which snake motions are optimal
for two retrograde traveling-wave body shapes---triangular and sinusoidal waves---across a wide range of frictional parameters and
incline angles. In the regime of large transverse friction coefficient, we find power-law
scalings for the optimal wave amplitudes and corresponding costs of locomotion. We give an asymptotic analysis to show that the optimal snake motions are traveling-wave motions
with amplitudes given by the same scaling laws found in the numerics.
\end{abstract}


\maketitle

\section{INTRODUCTION}
Snake locomotion has recently drawn interest from biologists, engineers and applied mathematicians \cite{Astley,Hu09, Maneewarn}.
A lack of limbs distinguishes snake kinematics from other common forms of animal locomotions such as swimming,
flying, and walking. Snakes propel themselves by a variety of gaits
including slithering, sidewinding, concertina motion, and rectilinear progression \cite{Hu09}.
They can move in terrestrial \cite{Gray,Jayne}, aquatic \cite{Shine}, and aerial \cite{Socha} environments.
Snake-like robots have shown the potential to realize similar locomotor ability, and have potential applications in
confined environments like narrow crevices \cite{Erkmen}, as well as rough terrain \cite{Yim}.
In such environments the ability to ascend an incline is fundamental, and various studies of snakes
and snake-like robots have been carried out on this subject.
Maneewarn and Maneechai \cite{Maneewarn} examined the pattern of crawling gaits in narrow inclined pipes with jointed modular snake robots
and found that high speeds were obtained with short-wavelength motions.
Hatton and Choset \cite{Hatton} focused on sidewinding gaits on inclines and presented stability conditions for snakes by
comparing sidewinding to a rolling elliptical trajectory and determining the minimum aspect ratio of the sidewinding pattern to
maintain stability.
Marvi and Hu \cite{Marvi} studied the concertina locomotion of snakes climbing on steep slopes and in vertical crevices.
They found that snakes can actively orient their scales and lift portions of their bodies to vary their frictional interactions with the surroundings.
Transeth \etal \cite{Transeth} considered the obstacle-aided locomotion of snake robots on inclined and vertical planes. Their model included both frictional forces and forces from rigid-body contacts with the obstacles, and their numerical results were consistent with their robotic studies. In this work, we focus on the slithering motion of snakes on an inclined plane by extending a recently-proposed
model for motions on a horizontal plane \cite{Hu09, Hu12, Jing, Alben} to motions on an incline.
Here the snake is a slender body whose curvature is prescribed as a function of arc length and time. For simplicity,
we do not consider elasticity or viscoelasticity in the snake body.
The external forces acting on the snake body are Coulomb friction with the ground, and gravity.
The model has shown good agreement with biological
snakes on a horizontal plane \cite{Hu09,Hu12}, and previous studies have used the model to find optimally efficient snake motions.
Hu and Shelley \cite{Hu12} prescribed a sinusoidal traveling-wave body motion and found the optimally efficient amplitude and wavelength of the traveling wave. Jing and Alben \cite{Jing} used the same model to consider the locomotion of two-
 and three-link snakes. They found optimal motions analytically and numerically in terms of the temporal function
for the internal angles between the links. Alben \cite{Alben} considered more general snake shapes and motions by
optimizing the curvature as a function of arc length and time with 45 (and 180) parameters, across the space of
frictional coefficients. He found that the optimal motions are two types of traveling-wave motions,
retrograde and direct waves,
for large and small transverse friction coefficients, respectively. In the large transverse friction coefficient regime,
he showed analytically that the optimal motion is a traveling wave, and found the scaling laws
for the wave amplitude and cost of locomotion with respect to the
friction coefficients, both numerically and analytically.

In this paper, we confine our discussion to the regime where the transverse friction coefficient is larger than the
tangential friction coefficients. This is the typical regime for biological and robotic snakes \cite{Hu12,Hirose}.
We prescribe the snake's motion as a retrograde
traveling wave with two shape profiles---triangular and sinusoidal---but with undetermined amplitudes.
The triangular wave motion has analytical solutions and embodies
many aspects of general traveling-wave motions. We obtain the optimal motions in terms of the amplitudes of the triangular wave in the
three-parameter space of transverse frictional coefficient, tangential (forward) frictional coefficient, and the incline angle.
We discuss the relative effects of these three parameters and find the upper
bound on the incline angle for the snake to maintain an upward motion.
We also find the power law scalings for the optimal amplitudes and corresponding costs of locomotion with respect to the three parameters.
For the sinusoidal wave motion, we use a numerical method to solve for the position of the snake body from its prescribed
curvature. We then obtain the optimal body shape numerically and show that it follows the same scaling laws as the triangular wave motion,
providing confirmation of those results.
In the last part of the paper, we analytically determine the asymptotic optima for more general snake motions in the
regime of large transverse friction coefficient. We obtain the same scalings analytically as were found for the
triangular and sinusoidal waves, which confirms and generalizes those results.
Our study of snake locomotion can also be generalized to other locomotor systems as long as the same frictional law applies. One example is the undulatory swimming of sandfish lizards in sand \cite{Hatton13,Ding}.

The paper is organized as follows: Section II describes the mathematical model for snake locomotion on an inclined plane.
Section III discusses the optima of the triangular wave motion, while Section IV studies the sinusoidal wave motion.
The analytical asymptotic discussion is presented in Section V, and the conclusions follow in Section VI.

\section{MODEL}
We use the same frictional snake model as \cite{Hu09,Hu12,Jing,Alben} to describe snake locomotion on an incline.
The position of the snake is given by $\bX(s,t)=(x(s,t),y(s,t))$,
a function of arc length $s$ and time $t$. The unit vectors tangent and normal to the snake body
are $\hs$ and $\hn$ respectively.
The snake is placed on a plane inclined at angle $\a$ with respect to the horizontal plane.
The $x$-$y$ axes are oriented with the $+x$ axis extending from origin directly up the incline and the
$+y$ axis rotated 90 degrees from it and directed across the incline. Height is constant along the
$y$ axis.
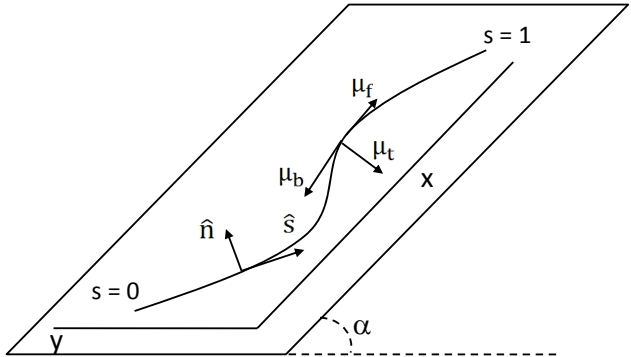
A schematic diagram is shown in figure \ref{fig:Schematic}.\Eonefigs{snakeschematic}{\label{fig:Schematic} Schematic of the snake position on a plane inclined at angle $\a$. The arc length $s$ is nondimensionalized by the snake length. The tangent and normal vectors are labeled at a point. Forward, backward and transverse velocities are shown with corresponding friction coefficients $\mu_f$, $\mu_b$ and $\mu_t$.}{3.5}

The tangent angle and the curvature are denoted $\th(s,t)$ and $\k(s,t)$. Given the curvature, the tangent angle and the position of the snake body
can be obtained by integrating:
\begin{eqnarray}
\th(s,t)&=&\th_0(t)+\int_0^s\k(s',t)\,ds',\\
x(s,t)&=&x_0(t)+\int_0^s\cos\th(s',t)\,ds',\\
y(s,t)&=&y_0(t)+\int_0^s\sin\th(s',t)\,ds'.
\end{eqnarray}
The trailing-edge position $(x_0,y_0)$ and the tangent angle $\th_0$ are determined by the force and torque balances for the snake:
\begin{eqnarray}
\int_0^L \rho\partial_{tt}xds&=&\int_0^Lf_x\,ds,\\
\int_0^L\rho\partial_{tt}yds&=&\int_0^Lf_y\,ds,\\
\int_0^L\rho\bX^{\perp}\cdot\partial_{tt}\bX \,ds&=&\int_0^L\bX^{\perp}\cdot\bF \,ds.
\end{eqnarray}
Here $\rho$ is the mass per unit length and $L$ is the length of the snake.
We assume the snake is locally inextensible, and $\rho$ and $L$ are constant in time.
$\bF$ is the external force per unit length acting on the snake. It includes two parts: the force due to Coulomb friction with the ground \cite{Hu09}, and gravity:
\begin{equation}
\mbox{$\bF = \rho g\cos{\a}\left[-\mu_t\left(\hptX\cdot\hn\right)\hn-\left(\mu_fH(\hptX\cdot\hs)+\mu_b(1-H(\hptX\cdot\hs))\right)\left(\hptX\cdot\hs\right)\hs\right]-\rho g G_\a.$}
\end{equation}
$H(\cdot)$ is the Heaviside function, and $G_\a=(\sin\a,0)^T$ represents the component of gravity in the downhill ($-x$) direction.
 The hats denote normalized vectors and we define $\hptX$ to be $0$ when the snake velocity is $0$.
The friction coefficients are $\mu_f,\mu_b$ and $\mu_t$ for motions in the forward $(\hs)$, backward $(-\hs)$ and transverse $(\pm\hn)$ directions. Without loss of generality we take
$\mu_f \leq \mu_b$, so the forward direction has the smaller
friction if it is unequal in the forward and backward directions.

We prescribe the curvature $\k(s,t)$ as a time-periodic function with period $T$ and nondimensionalize equations
(4)-(6) by length $L$, time $T$ and mass $\rho L$. We then obtain:
\bqa
\frac{L}{gT^2}\int_0^1\partial_{tt}x\,ds&=&\int_0^1f_x\,ds,\\
\frac{L}{gT^2}\int_0^1\partial_{tt}y\,ds&=&\int_0^1f_y\,ds,\\
\frac{L}{gT^2}\int_0^1\bX^{\perp}\cdot\partial_{tt}\bX \,ds&=&\int_0^L\bX^{\perp}\cdot\bF \,ds.
\eqa
We neglect the snake's inertia for simplicity, as $L/gT^2 \ll 1$ for typical steady snake locomotion
\cite{Hu09} and set the left hand sides of (8)-(10) to zero. As discussed by Alben \cite{Alben},
the model still maintains a good representation of real snake motions. We then obtain the following dimensionless force and torque equations:
\begin{eqnarray}
\int_0^Lf_xds = 0,\\
\int_0^Lf_yds = 0,\\
\int_0^L\bX^{\perp}\cdot \bF ds = 0.
\end{eqnarray}
and the dimensionless force $\bF$ becomes:
\begin{equation}
\bF = \cos{\a}\left[-\mu_t\left(\hptX\cdot\hn\right)\hn-\left(\mu_f H(\hptX\cdot\hs)+\mu_b(1-H(\hptX\cdot\hs))\right)\left(\hptX\cdot\hs\right)\hs\right]-G_\a.
\end{equation}

The frictional force tends to 0 as $\a$ approaches $\pi/2$. On a strictly vertically
plane, frictional force is unable to balance gravity, so planar locomotion is not obtained
by our model in this case (however, snakes can ascend vertical crevices
in a non-planar concertina motion \cite{Marvi}).

Given the curvature $\k(s,t)$, we solve the three nonlinear equations (11)-(13) at each time $t$ for the three unknowns $x_0(t), y_0(t)$ and $\theta_0(t)$.
Then we obtain the snake's position as a function of time by equations (1)-(3). We define the cost of locomotion as
\bq
\eta=\frac{W}{d},
\eq
where $d$ is the distance traveled by the snake's center of mass over one period
\bq
d=\sqrt{\left(\int_0^1x(s,1)-x(s,0)\,ds\right)^2+\left(\int_0^1y(s,1)-y(s,0)\,ds\right)^2},
\eq
 and $W$ is the work done by the snake against frictional forces and gravity over one period
 \bq
 W=\int_0^1\int_0^1 -\bF(s,t)\cdot\ptX(s,t)\,ds\,dt.
 \eq
Our objective is to find the curvature $\k(s,t)$ that minimizes $\eta$. We choose the initial
orientation of the snake so that its center of mass travels only in the $x$ direction (up
the incline) over one period of motion.

We briefly mention the case in which the snake moves down the incline,
which is equivalent to setting $\a<0$. In this case the snake can slide down the incline
with no change of shape, and the work done by gravity and friction are equal
in magnitude and opposite in sign. Thus the cost of locomotion is 0 regardless of the frictional parameters.
 A straight body with $y(s,t)=0$ experiences purely tangential friction, and achieves the fastest speed among possible
body shapes.

By contrast, when the snake ascends the incline, i.e. $\a\geq 0$, we will consider
traveling wave motions in which net tangential friction and gravity both act in the $-x$ direction and transverse friction is necessary to balance the $x$-force equation.
 In the following discussions we only consider
$\a\geq 0$, and look for nontrivial $\k(s,t)$ to minimize the cost of locomotion.

\section{TRIANGULAR WAVE BODY SHAPE}
\subsection{Range of $\a$}
We start our discussion with a triangular wave body motion. It was studied on a level plane in \cite{Alben},
and the snake's position, angle, velocity and work can all be obtained analytically. The motion is useful
to consider because it embodies
many aspects of more general traveling wave motions, while the shape dynamics are easy to understand.

The triangular wave has zero mean $y$ deflection:
\bq
y(s,t)=A\dashint\sgn(\sin(2\pi(s+t)))\,ds,
\eq
with the unit tangent and normal vectors:
\bq
\hs=\left(\begin{array}{c}
\sqrt{1-A^2}\\
A\sgn(\sin(2\pi(s+t)))\end{array}\right),\quad
\hn=\left(\begin{array}{c}
-A\sgn(\sin(2\pi(s+t)))\\
\sqrt{1-A^2}.\end{array}\right).
\eq
The force and torque balance equations are satisfied when the snake moves forward
with a constant speed $U$. Then the horizontal and vertical speeds are
\bq
\partial_tx(s,t)=U,\ \partial_ty(s,t)=A\sgn(\sin(2\pi(s+t))).
\eq
The net $y$-force and torque for such a motion are identically zero.
We determine the horizontal speed $U$ by the $x$-force balance equation:
\bq
\int\cos{\a}\left(-\mu_t\hptX\cdot\hn n_x-\mf\hptX\cdot\hs s_x\right)-\sin{\a}\,ds = 0
\eq
Since $\mu_f \leq \mu_b$, and the tangential velocity is uniformly forward or backward
over the whole snake body for the triangular wave, the most efficient motion is obtained when
the snake moves forward. Thus $\mu_b$ does not appear in the frictional force in (21). Notice that the tangential frictional force and gravity both have a
component in the $-x$ direction, and the transverse frictional force is the only source for a balancing force in the $+x$ direction, up the incline.
Solving (21) for $U$ we obtain:
\bq
U=\frac{\displaystyle \left(A^4\left(\frac{\mt}{\mf}-1\right)^2+A^2\left(\mtmf-1\right)\right)\sqrt{1-A^2}-A\frac{\tan\a}{\mf} \sqrt{1+A^2\left(\frac{\mt^2}{\mf^2}-1\right)-\frac{\tan^2\a}{\mf^2}}}{\displaystyle\left(1+A^2\left(\mtmf-1\right)\right)^2-\frac{\tan^2\a}{\mf^2}}.
\eq
The speed of the snake is a function of $A$, $\mu_t$ , $\mf$ and $\a$. We require the speed to be real and nonnegative,
and therefore the incline angle $\a$ must satisfy the following inequality:
\bq
\a\leq \arctan(A(\mt-\mf)\sqrt{1-A^2}).
\eq

Here we use the fact that the amplitude $A \leq 1$ in the triangular wave model and $\mu_t>\mf$. In figure \ref{fig:Alpharegion},
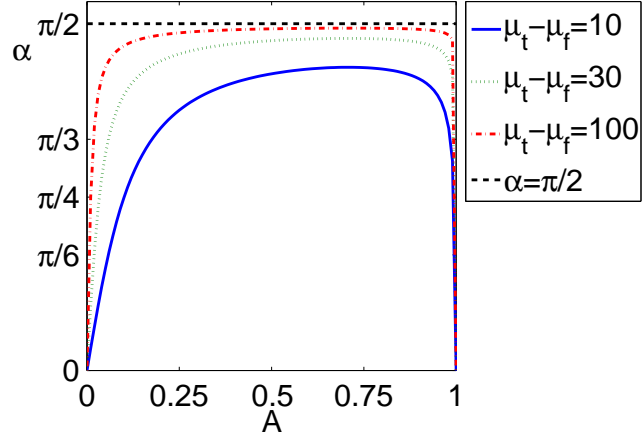
we plot the upper bound of $\a$ according to inequality (23). \Eonefigs{alpharegion1}{\label{fig:Alpharegion} Lines showing the boundaries of the region on nonnegative forward velocity $U$ in the space of
$A$ and $\a$, for different $\mu_t-\mf = 10, 30$
and $100$. On the lines, $U$ = 0, and below the lines, $U > 0$.
The black dashed line shows $\a=\pi/2$, giving a vertical incline.} {3.5}
For a given value of $\mt - \mf$, nonnegative speed is obtained for $\a$
in the region bounded by a curved line (labeled by $\mt - \mf$) and the horizontal ($A$) axis.
As $\mt$ increases,
larger transverse friction can be generated for the same $A$, and therefore the range of $\a$ increases accordingly. As $\mf$ becomes larger,
the tangential motion produces a stronger downhill drag which inhibits upward motion, so the corresponding $\a$ range decreases.
When the amplitude $A$ varies from 0 to 1, both the transverse and tangential frictional forces vary and their $x$-components have
opposite sign. As a result, the range of $\a$ is non-monotonic with respect to $A$.
The largest upper bound is obtained at $A=\sqrt{2}/2$.

\subsection{Optima and other results}
In the triangular wave motion, the velocity and power are both constant over time,
so we can simplify the cost of locomotion as
\bq
\eta=\frac{W}{d}=\frac{P}{U}=\frac{\displaystyle\int_0^1 -\bF\cdot\ptX ds}{U},
\eq
and obtain
\bq
\eta=\frac{\cos\a}{\sqrt{U^2+A^2}}\left(A^2(\mt-\mf)\left(U-2\sqrt{1-A^2}-\frac{A^2}{U}\right)+\mf U+\frac{\mt A^2}{U}\right)+\sin\a.
\eq
We plug the value of $U$ (22) into (25) and minimize $\eta$ with respect to $A$. Then in the large-$\mt$ limit we obtain
the optimal cost of locomotion and corresponding amplitude as:
\bqa
\min \eta\longrightarrow(\mf\cos\a+\sin\a)\left(1+\left(\frac{2\mf}{\mt}\right)^{1/2}\right),\\
A\longrightarrow 2^{1/4}\mu_t^{-1/4}\left(\mf^{1/2}+\frac{\tan\a}{\mf^{1/2}}\right)^{1/2}.
\eqa

We plot the optimal motion with $\a=\pi/4$, $\mf=1$ and $\mt=10$ over one period in figure \ref{fig:triangleshape}.
We manually offset the body by a constant increment in the $-y$ direction with every snapshot to clearly show the
individual bodies, but we note that for
the triangular wave motion, the snake's center of mass moves purely along the $x$-axis.
The peak of the snake shifts to the left in the figure, which indicates
the snake moves slower than the traveling triangular wave. The snake slips transversely in the $-x$ direction to obtain a thrust force in the $+x$ direction that balances the drag forces due to
tangential friction and gravity.
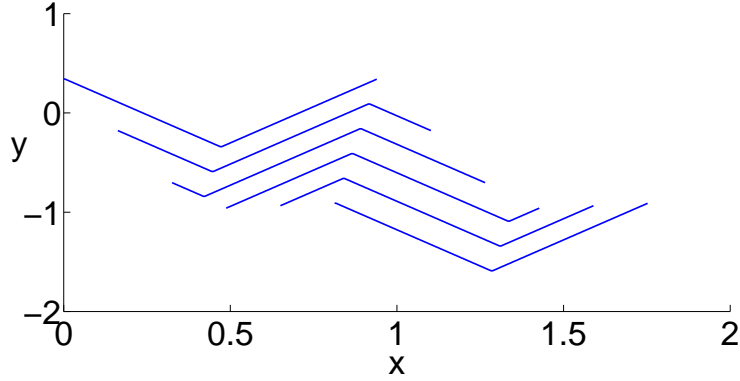
\Eonefigs{triangularshape}{\label{fig:triangleshape}The optimal snake trajectory of the triangular
wave body shape over one period obtained with $\a=\pi/4$, $\mf=1$ and $\mt=10$.}{4.5}

In figure \ref{fig:trianglewave} we plot the optimal $A$ and $\eta$ with respect to $\a$, $\mu_t$, and $\mf$.
Our results extend to $\mu_t$ below the large-$\mu_t$ limit, but in this limit the results agree with (26) and (27). For each parameter set, we minimize $\eta$ over $A$ using equations (22) and (25).
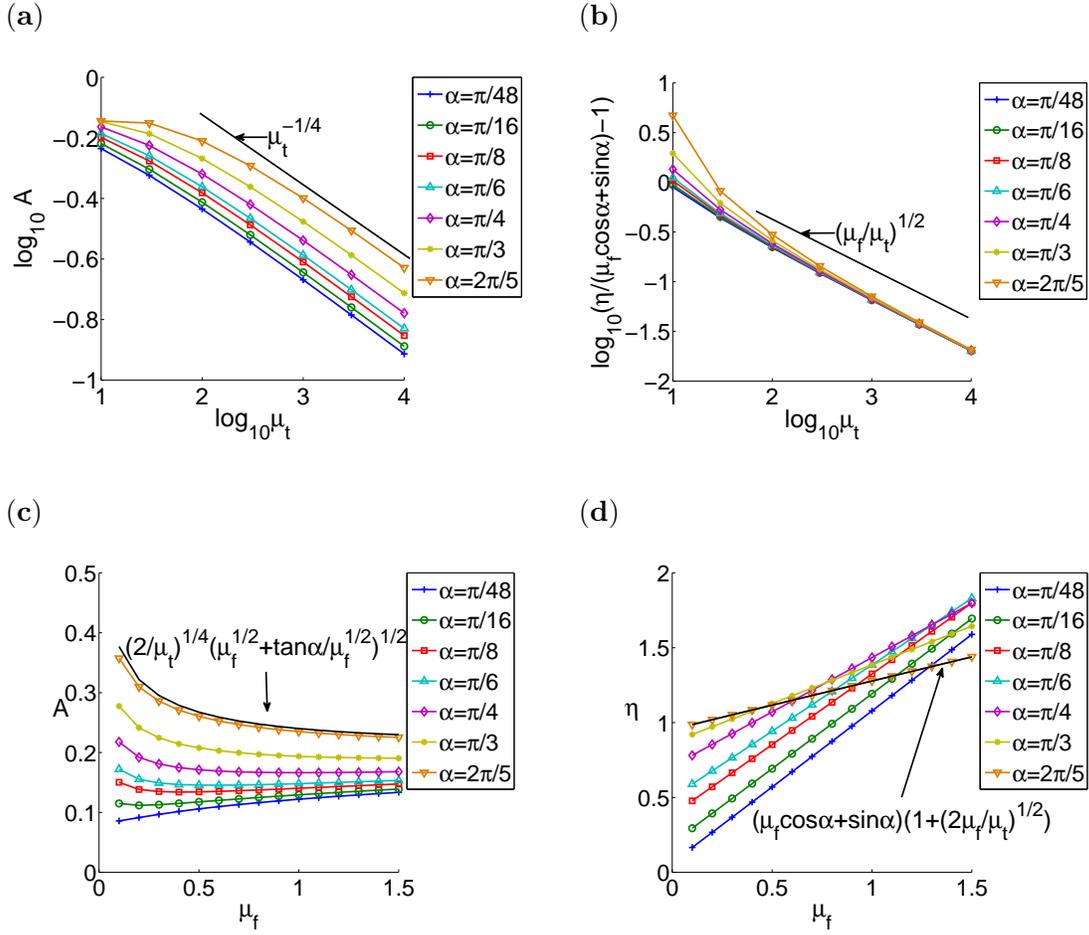
\Efourfigs{TriangleWaveAvsMu}{TriangleWaveEtavsMu}{TriangleWaveAvsMuf2}{TriangleWaveEtavsMuf1}{\label{fig:trianglewave} Scaling
laws of the triangle wave optima. (a) $\log_{10} A$ vs. $ \log_{10}\mu_t$ with various $\a$ and fixed $\mf=1$. The solid line indicates
the scaling $\mu_t^{-1/4}$. (b) $\log_{10} (\eta/(\mf\cos\a+\sin\a)-1)$ vs. $ \log_{10}\mu_t$ with various $\a$ and fixed $\mf=1$.
The solid line shows the scaling $(\mf/\mu_t)^{1/2}$. (c) $A$ vs. $\mf$ with various $\alpha$ and fixed $\mt=10000$.
The solid line is the asymptotic solution $A=2^{1/4}\mu_t^{-1/4}(\mf^{1/2}+\tan\a/\mf^{1/2})^{1/2}$, obtained with $\a=2\pi/5$.
(d) $\eta$ vs. $\mf$ with various $\a$ and fixed $\mt=10000$. The solid line shows the asymptotic optima $\eta=(\mf\cos\a+\sin\a)(1+(2\mf/\mt)^{1/2})$ with
 $\a=2\pi/5$.}{2.8}{2.8}

We plot $A$ versus $\mt$ in figure \ref{fig:trianglewave}a
with fixed $\mf=1$ and vary the parameter $\a$. The asymptotic scaling of $\mt^{-1/4}$ is shown with the solid line.
The corresponding $\eta$ and the scaling $(\mf/\mt)^{1/2}$ (solid line) are plotted in figure \ref{fig:trianglewave}b.
The optimal magnitude $A$ and cost of locomotion $\eta$ both decrease with larger $\mt$.
We vary $\mf$ and $\a$ in figure \ref{fig:trianglewave}c and d
 with fixed $\mt=10000$, and plot the optimal $A$ versus $\mf$ and $\eta$ versus $\mf$ respectively.
The analytical solutions of (26) and (27) for $\a=2\pi/5$ are shown with solid lines in both figures,
 and they agree well with the numerical results at the largest $\mu_t$ (downward-pointing triangles).
The optimal amplitude $A$ achieves its minimum at $\mf=\tan\a$, while the cost of locomotion $\eta$ monotonically increases
with $\mf$ as $\partial_{\mf}\eta > 0$. When $\a$ goes up, the optimal $A$ increases accordingly and its
minimum over $\mf$ shifts to larger $\mf$ (figure \ref{fig:trianglewave}c).
But the cost of locomotion varies non-monotonically with $\a$. We can rewrite (27) as
\bq
\min\eta\longrightarrow \sqrt{\mf^2+1}\sin\left(\a+\arcsin \frac{\mf}{\sqrt{\mf^2+1}}\right)\left(1+\left(\frac{2\mf}{\mt}\right)^{1/2}\right),
\eq
The least efficient optimum is obtained when
\bq
\a^*=\frac{\pi}{2}-\arcsin\frac{\mf}{\sqrt{\mf^2+1}}.
\eq
We call $\a^*$ the critical incline angle. The optimal snake moves more efficiently
when the incline is either shallower or steeper than the incline at the critical angle.
The critical incline angle $\a^*$ only depends on the tangential friction coefficient and
becomes smaller as $\mf$ increases. On a steeper slope, more work is done against gravity and less work against forward friction, for a given distance travelled. Thus when $\mu_f$ increases,
efficiency can be improved by making the slope steeper (and adjusting the amplitude
to achieve the optimum at the new set of parameters).

To better understand the effects of the parameters $\mt,\mf$, and $\a$, we plot the costs of locomotion due to transverse friction alone
and tangential friction alone versus $A$ in figure \ref{fig:friction}. We decompose the cost of locomotion (25) into three parts:
\bq
\eta=\eta_t+\eta_f+\eta_g,
\eq
where
\bqa
\eta_t&=&\frac{\cos\a}{\sqrt{U^2+A^2}}\left(UA^2\mt-2A^2\sqrt{1-A^2}\mt-\frac{A^4\mt}{U}+\frac{A^2\mt}{U}\right),\\
\eta_f&=&\frac{\cos\a}{\sqrt{U^2+A^2}}\left(U\mf(1-A^2)+2A^2\sqrt{1-A^2}\mf+\frac{A^4\mf}{U}\right),\\
\eta_g&=&\sin\a
\eqa
are the costs due to transverse friction, forward tangential friction, and gravity, respectively.

In figure \ref{fig:friction}, we vary one of the parameters $\mt, \mf$, and $\a$ in turn and keep the other two fixed.
We use solid lines for transverse friction and dashed lines for tangential friction in all panels. In general, as the amplitude $A$ becomes larger,
the cost due to transverse friction decreases while the cost due to tangential friction increases.
 The optimal $\eta$ is obtained when the slopes of the two costs are equal in magnitude and opposite in sign. $\eta_g$ is
independent of $A$ so it does not play a role in determining the optimal amplitude with respect to $A$.
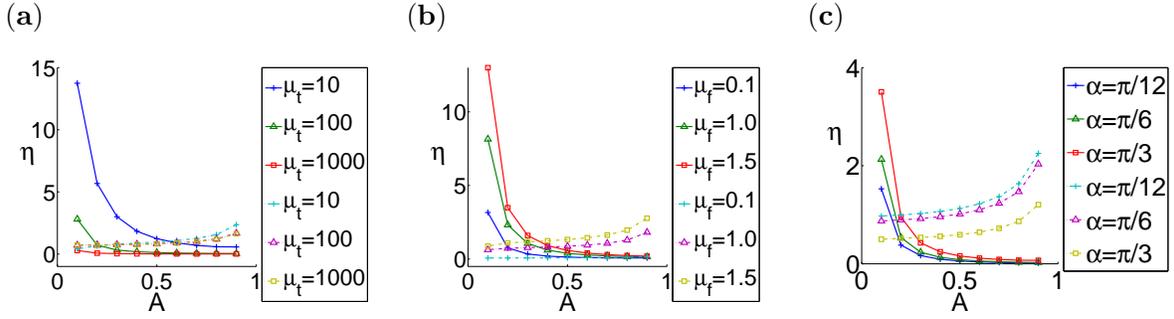
\Ethreefigs{frictionforce2}{frictionforce3}{frictionforce}{\label{fig:friction} Cost of locomotion versus $A$ with various $\mt$,
 $\mf$ and $\a$. (a) $\eta$ vs. $A$ with various $\mt$ and
fixed $\a=\pi/4$ and $\mf=1$. (b) $\eta$ vs. $A$ with various $\mf$ and fixed $\a=\pi/4$ and $\mt=30$. (c) $\eta$ vs. $A$ with various $\a$ and fixed $\mt=100$ and $\mf=1$. The solid lines show the cost due to
transverse friction and the dashed lines denote the cost due to tangential friction.}{1.9}{1.9}{1.9}

In figure \ref{fig:friction}a, the sums of the costs of the frictional forces become smaller as $\mt$ increases. Thus the optimal $\eta$ decreases as well
as shown in figure \ref{fig:trianglewave}b. For a given motion (a given A), the slope of the tangential cost is almost unchanged as $\mt$ goes up,
while the magnitude of the slope of the transverse cost quickly decays. The point where the two slopes balance shifts to the left at larger $\mt$. The optimal motion is thus obtained at a smaller amplitude as $\mt$ increases.
 We show the results only for $\a=\pi/4$ and $\mf=1$ in the figure panel, but the same phenomenon holds for all $\a$ and $\mf$.
 Physically, as the transverse coefficient increases,
the snake can obtain the same amount of forward force from transverse friction
with less deflection of the body and less slipping in the transverse direction, and the cost of the
tangential friction is reduced as well
due to a shorter path travelled. Thus, the total cost of locomotion $\eta$ decreases with $\mt$.

We show in figure \ref{fig:friction}b that the costs due to transverse friction and tangential friction both increase as $\mf$ increases.
When the friction coefficient $\mf$ is larger, the snake of the same deflection experiences a stronger downward drag caused by tangential friction, increasing the slipping and consequently the work done against transverse friction as well.
The optimal amplitude $A$ varies non-monotonically with $\mf$ as shown in figure \ref{fig:trianglewave}c. The slopes of both costs increase with $\mf$ for given $A$. When $\mf<\tan\a$, the point where the two slopes are equal in magnitude shifts to smaller $A$ as $\mf$ increases. When $\mf>\tan\a$, the balanced point shifts to larger $A$.

In figure \ref{fig:friction}c, we fix $\mt=100$ and $\mf=1$, and vary the incline angle $\a$. The cost due to gravity is $\sin\a$ for the triangular
wave motion and thus always increases with $\a$. Meanwhile, the tangential cost decreases with $\a$ while the transverse cost increases.
The competition of these three costs makes $\eta$ non-monotonic with $\a$ as shown in figure \ref{fig:trianglewave}d. For a given motion,
the slope of the tangential cost with respect to $A$ is nearly unchanged as the incline becomes steeper. However, the magnitude of the slope of the
transverse cost increases with $\a$, so a point with a given slope of the transverse cost shifts to larger $A$ as
$\a$ increases. Therefore, the point where two slopes are equal and opposite shifts to the right as $\a$ goes up, and the optimal $A$ in figure \ref{fig:trianglewave}a
and c grows with $\a$.

\section{SINUSOIDAL BODY SHAPE }
\subsection{Numerical Method}
We now consider an alternative, sinusoidal snake motion, to check the dependence of our results on the snake shape.
We again determine the snake shape which minimizes $\eta$ for a given parameter
set $(\mu_t,\mu_f,\mu_b,\a)$. We consider the sinusoidal body shape
\bq
\k(s,t)=K\cos(n\pi s+2\pi t)
\eq
where the curvature is prescribed as a sinusoidal function with $t$-period $1$. We fix the wave number $n$ in this work
and look for the optimal constant $K$ to minimize $\eta$ for a given $(\mu_t,\mf,\mu_b,\a)$. Later we show that although the wave
number $n$ affects the optimal value of $K$, it does not change the dependence of the optimal $K$ on the other parameters.
We fix $n=6$ in this section, since for a horizontal plane, the lowest cost of locomotion is
obtained in the limit of large wave numbers according to \cite{Alben}, and $n=6$ approximates
this limit while only requiring a moderate number of grid points in arc length
along the snake to discretize the equations accurately. We find the optimal $\eta$ on a sequence of one-dimensional meshes of $K$ values with decreasing spacing.
Each mesh in the sequence is centered near the minimizer from the previous coarser mesh. The fourth mesh used has a
mesh size of $10^{-3}$. We then use a quadratic curve to fit the data around the minimum point on the fourth mesh and obtain the optimal $K$
based on a final, fifth mesh, with mesh width $10^{-6}$.
In \cite{Alben} Alben used a BFGS algorithm to minimize the cost of locomotion when the curvature is described by a double-series expansion with 45 parameters. In this work we optimize over only a single parameter (the amplitude), for a given shape.
We find that a direct search over the parameter space is typically fast enough since $\eta$ varies smoothly with $K$ in the
regime of physically admissable, non-overlapping shapes.

The algorithm requires fast routines to evaluate $\eta$. Here we describe our numerical scheme that solves
for the work, distance and cost of locomotion. Given the curvature $\k(s,t)$, we solve the three nonlinear equations
(11)-(13) at each time step, over a period, for the three unknowns $x_0(t), y_0(t)$ and $\th_0(t)$.
 Then we use equations (1)-(3) to compute $x(t), y(t), \th(t)$ and obtain $d, W,$ and $\eta$ over one period.

We discretize the period interval uniformly with $m$ time points: $\{0,1/m,\cdots,1-1/m\}$. At each time, we rewrite equations (11)-(13)
as equations in unknowns $\{\partial_t x_0, \partial_t y_0, \partial_t\th_0\}$ by taking time derivatives on both sides. We solve for $\{\ptx0, \pty0,\ptth0\}$
over one period and then integrate to obtain $\{x_0,y_0,\th_0\}$. The advantage of replacing $x_0,y_0$ and $\th_0$ with their time derivatives as
 variables is that it can reduce the numerical error in computing the discrete time derivatives and decrease the computational
complexity by decoupling the large system of $3m$ equations in $3m$ unknowns into $m$ decoupled systems each containing only $3$ equations in $3$ unknowns \cite{Alben}.

We design a time-marching scheme which is second-order in both time and space to solve for $\{\partial_t x_0, \partial_t y_0,\partial_t\th_0\}$ at each time level.
At time $t=0$, we set $\{x_0^0,y_0^0,\th_0^0\}$ to $\{0,0,0\}$ and solve for $\{\partial_t x_0^0,\partial_t y_0^0\,\partial_t\th_0^0\}$ using Newton's method
with a finite-difference Jacobian matrix as described in \cite{Alben}. At each time level $i$, we need $\{x_0^i,y_0^i,\th_0^i\}$ to carry out the computation. We obtain an initial guess for the current position and
angle by a forward Euler scheme using the previous step solutions $\{\ptx0^{i-1},\pty0^{i-1},\ptth0^{i-1}\}$ and $\{x_0^{i-1},y_0^{i-1},\th_0^{i-1}\}$.
 We then solve for $\{\ptx0^{i},\pty0^{i},\ptth0^{i}\}$ at time $i$. Then we correct $\{x_0^{i},y_0^{i},\th_0^{i}\}$ by
integrating $\{\ptx0,\pty0,\ptth0\}$ from $t=0$ to $t=i/m$. Our method is similar to the prediction-correction
algorithm for solving a system of ODEs. We can iterate the same procedure until
a certain accuracy is obtained. We
find that second-order temporal accuracy is achieved by only
performing the above procedure once.

\subsection{Optima and other results}


We first consider the parameter regime where $\mt\gg\mf$. In this region, the tangential motion is purely in the forward direction for the sinusoidal
 wave motion. Therefore, as for the triangular wave, the $\mu_b$ term drops out of the force law,
and the parameter space is reduced to $\{\mt,\mf,\a\}$. We plot the optimal snake trajectory with parameters $\a=\pi/4$, $\mf=1$ and $\mt=10$
over one period in figure \ref{fig:travelshape}. The snake moves from left to right and its center of mass moves
mainly along the $x$ direction.
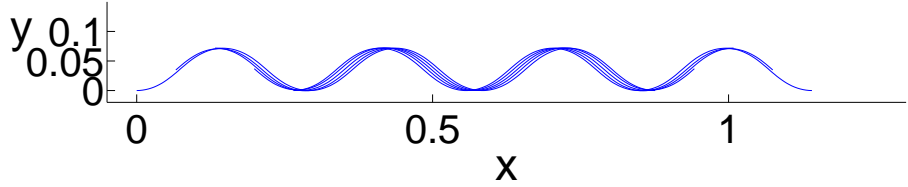
 \Eonefigs{travelshape1}{\label{fig:travelshape}The optimal snake trajectory of the sinusoidal wave body shape over one period obtained
with $\a=\pi/4$, $\mf=1$ and $\mt=10$.}{5}

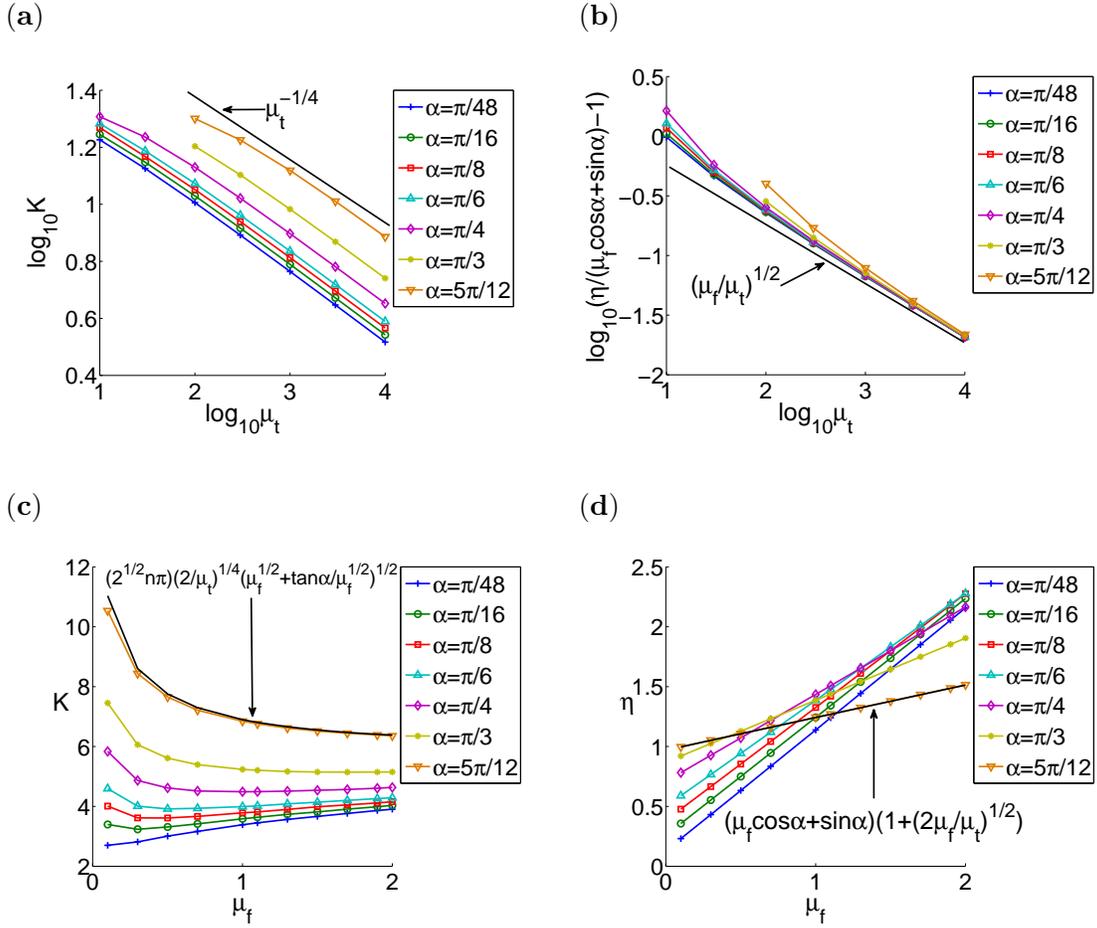
\Efourfigs{TravelWaveAvsMu}{TravelWaveEtavsMu}{TravelWaveAvsMuf2}{TravelWaveEtavsMuf2}{\label{fig:Travel} Scaling laws for the sinusoidal wave optima.
 (a) $\log_{10} K$ vs. $ \log_{10}\mu_t$ with various $\a$ and fixed $\mf=1$. The solid line shows the scaling law $\mu_t^{-1/4}$.
(b) $\log_{10} (\eta/(\mf\cos\a+\sin\a)-1)$ vs. $ \log_{10}\mu_t$ with various $\a$ and fixed $\mf=1$.
The solid line denotes the scaling $(\mf/\mu_t)^{1/2}$. (c) $K$ vs. $\mf$ with various $\alpha$ and fixed $\mt=10000$. The solid line is
the analytical optimum $K=\sqrt{2}n\pi(2/\mt)^{1/4}(\mf^{1/2}+\tan\a/\mf^{1/2})^{1/2}$ obtained with $\a=5\pi/12.$
(d) $\eta$ vs. $\mf$ with various $\a$ and fixed $\mt=10000$. The solid line is the graph of the optimal
$\eta = (\mf\cos\a+\sin\a)(1+(2\mf/\mt)^{1/2})$ with $\a=5\pi/12.$}{2.8}{2.8}
In figure \ref{fig:Travel}, we vary $\mt$, $\mf$ and $\a$, and plot the optimal $K$ and cost of locomotion $\eta$ versus these parameters respectively.
Some data points are ignored because there is no solution with non-negative $x$-velocity for the corresponding parameter values.
We fix $\mf=1$ and plot the optimal $K$ and $\eta$ versus $\mt$ with various $\a$ in figure \ref{fig:Travel}a and b.
In figure \ref{fig:Travel}c and d, we fix $\mt=10000$ and vary $\mf$ from $0.1$ to $2$ with different $\a$.
We find that the optima for the sinusoidal wave motion satisfy essentially
the same scaling laws as the triangular wave motion in (26)
and (27). The cost of locomotion is the same as (26) and the amplitude $K$ is scaled by an extra factor $\sqrt{2}n\pi$ yielding a deflection amplitude that agrees with (27).

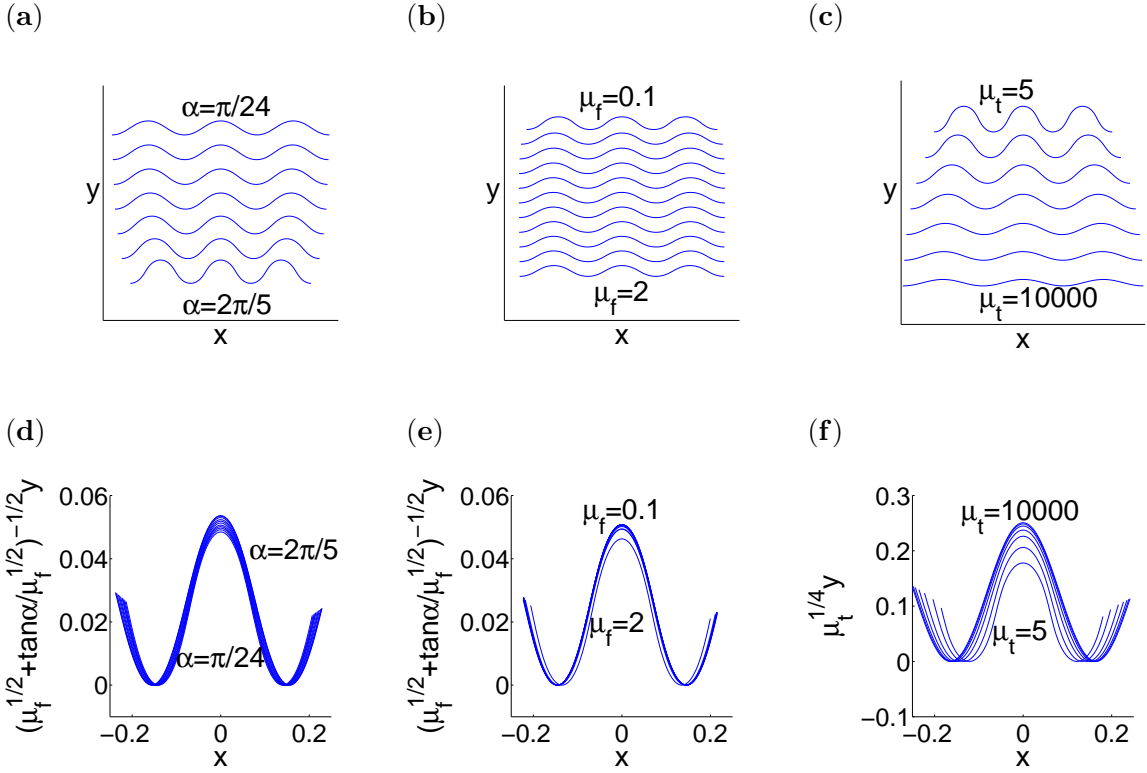
\ESixfigs{travelshapealpha}{travelshapemuf}{travelshapemut}{optimalrescalealpha}{optimalrescalemuf}{optimalrescalemut} {\label{fig:traveloptima}Optimal
 snake shapes at the instant $t=0$ for (a) various $\alpha$ from $\pi/24$ to $2\pi/5$ and fixed $\mf=1$ and $\mt=100$; (b) various $\mf$ from $0.1$ to $2$
 and fixed $\a=\pi/4$ and $\mt=100$; and (c) various $\mt$ from $5$ to $10000$ and fixed $\a=\pi/4$ and $\mf=1$.
(d) The deflection of the body rescaled by $(\mf^{1/2}+\tan\a/\mf^{1/2})^{-1/2}$ corresponds to (a).
(e) The deflection rescaled by $(\mf^{1/2}+\tan\a/\mf^{1/2})^{-1/2}$ for (b). (f) The deflection rescaled by $\mu_t^{1/4}$ with the same parameters as (c).}{1.9}

In figure \ref{fig:traveloptima}, we show the optimal snake body shapes corresponding to different $\a$, $\mf$ and $\mt$ in panels a, b
and c respectively. We displace the different bodies vertically so that they are easier to distinguish.
In figure \ref{fig:traveloptima}d,
e, and f, we rescale the deflections of the optimal shapes with $\a$, $\mf$, and $\mt$ according to the scaling law for the optimal amplitude (27).
The centers of mass for all bodies are located at the origin. Here we zoom in on the portion of the body nearest to the center.
We find a good collapse of the bodies after rescaling.

In the regime where $\mt$ and $\mf$ are comparable, the sinusoidal body shape model may not yield forward motion.
For snake locomotion on a level plane,
Alben \cite{Alben} found that when $\mt/\mf \lesssim 6$, the optimal snake shape is no longer a retrograde traveling wave.
He identified a set of locally optimal motions and classified some
of them as racheting motions. When the snake climbs uphill and $\mt$ is small, the traveling wave may not be able to provide enough uphill thrust
to balance gravity and the drag due to forward motion.
The snake may therefore use locomotion modes other than slithering to maintain its position on the incline.
For example, concertina locomotion is often observed for snakes moving inside an inclined tunnel. Marvi and Hu \cite{Marvi} found that some snakes can resist
 sliding by lifting part of the body and reducing the contact with the ground to
several localized regions.
This shows that a planar model is not sufficient to describe snake locomotion on inclines
when the ratio of the transverse-to-tangential friction coefficient is small.
Our future work may consider three-dimensional motions to better understand this interesting parameter regime.
\section{Asymptotic Analysis}
\subsection{Optimal Shape Dynamics}
We now analytically determine how the optimal snake motion depends on the parameters $\{\mt,\mf,\a\}$, thus providing theoretical
confirmation of our previous results and extending them to general shapes, in the large-$\mu_t$ regime.
We assume
that the mean direction of the motion is aligned with the $x$-axis and the deflections along the $y$ direction are small, i.e. $|y|$ and all of
its temporal and spatial derivatives $|\partial_ty|,|\partial_sy|,|\partial_t^2y|,\ldots,$ are $O(\mu_t^\beta)$ for some negative $\beta$.
We also assume that the tangential motion is only in the forward direction, to simplify the derivation.
We first expand each term in the force and torque balance equations in powers of $|y|$ and retain only the terms which are dominant at large $\mu_t$.
 A detailed discussion of the expansion of each term can be found in \cite{Alben}.  If we only keep the lowest powers in $y$ from each expression,
the $x$-force balance equation becomes:
\bq
\int_0^1-\cos\a\left(\mf+\mu_t\partial_sy\left(\partial_sy-\frac{\partial_ty}{U}\right)\right)-\sin\a\, ds=0.
\eq
where $U(t)\equiv \overline{\partial_tx(s,t)}$ is the $s$-averaged horizontal velocity.
The three terms in the integral represent the drag due to forward friction,
the thrust due to transverse friction, and gravity, respectively. We note that both
tangential friction and gravity forces act in the
$-x$ direction, and transverse friction is essential to maintain the $x$-force balance.
In \cite{Alben} it is shown that for large $\mu_t$, a minimizer of the cost of locomotion
should approximate a traveling wave motion. Therefore we pose the shape dynamics as
\bq
y(s,t)=g(s+U_wt),
\eq
which is a traveling wave with a prescribed wave speed $U_w$. $U_w$ is different from $U(t)$ in general,
for otherwise the snake moves purely tangentially with no transverse motion. Here $g(s+U_wt)$ is a periodic function with period $U_w$.
We obtain an equation for $U$ in terms of $U_w$ and $g$ by substituting (36) into (35):
\bq
-\mf\cos\a-\mt\cos\a\left(1-\frac{U_w}{U}\right)\langle g'(s+U_wt)^2\rangle =\sin\a,
\eq
where $\langle g'(s+U_wt)^2\rangle\equiv\int_0^1g'(s+U_wt)^2\,ds$. As $\a\rightarrow\pi/2$, $\cos\a\rightarrow 0$ and no frictional force occurs.
Therefore, in the limit $\a\rightarrow\pi/2$, we require $\mu_t\rightarrow\infty$ such that $\mt\cos\a\rightarrow\infty$, i.e, the speed at
which $\a$ tends to $\pi/2$ depends on the speed at which $\mu_t$ tends to infinity. This requirement is similar
to the upper bound of $\a$ in the triangular wave motion to obtain forward motion. As $\mu_t\cos\a\rightarrow\infty$, equation (37) holds with $(1-U_w/U)\rightarrow0^{-}$.
The traveling wave moves backward along the snake at speed $U_w$ while the snake moves forward at a speed $U$ slightly
less than $U_w$.
Therefore, the snake slips transversely to itself, which provides an uphill thrust to balance
gravity and the drag due to tangential friction.

In \cite{Alben} it is shown that one must expand the terms in the force balance equation in higher powers of $|y|$
to obtain an optimal motion. We do so, again assuming $y=g(s+U_wt)$.
Then the force balance equation becomes:
\bq
\cos\a\left(\mf+\mu_t\left[\left(1-\frac{U_w}{U}\right)\langle g'(s+U_wt)^2\rangle+\frac{1}{2}\langle g'(s+U_wt)^2\rangle^2\right]\right)=-\sin\a,
\eq
and the cost of locomotion is:
\bq
\eta \sim \sin\a+\int_0^1\cos\a\mf(1+\frac{1}{2}\gpavgs)+\cos\a\mu_t\left(1-\frac{U_w}{U}+\frac{1}{2}\gpavgs\right)^2\gpavgs +O(\mf|g|^4,\mu_t|g|^8)\, dt.
\eq
Equation (39) is shown in a simplified form, using the result that $(1-U_w/U)\rightarrow 0$ as $\mt\cos\a\rightarrow\infty$.
We substitute equation (38) into (39) and obtain:
\bq
\eta=1/\int_0^1\frac{1}{\displaystyle \sin\a+\cos\a\mf(1+\frac{1}{2}\gpavgs)+\frac{\cos\a(\mf+\tan\a)^2}{\mu_t\gpavgs}+O(\mf|g|^4,\mu_t|g|^8)} dt.
\eq
If we approximate $\gpavg$ as constant in time, we obtain
\bq
\eta=\mf\cos\a+\sin\a+\frac{\mf\cos\a}{2}\gpavgs+\frac{\cos\a(\mf+\tan\a)^2}{\mu_t\gpavgs}+O(\mf|g|^4,\mu_t|g|^8),
\eq
which is minimized for
\bq
\gpavgs^{1/2}=2^{1/4}\mu_t^{-1/4}\left(\mf^{1/2}+\frac{\tan\a}{\mf^{1/2}}\right)^{1/2},
\eq
and the corresponding optimal cost of locomotion is :
 \bq
 \eta=(\mf\cos\a+\sin\a)\left(1+\left(\frac{2\mf}{\mt}\right)^{1/2}\right).
\eq

In the triangular wave motion, we have
\bq
y(s,t)=g(s+t)=A\dashint\sgn(\sin(2\pi(s+t)))ds.
\eq

By using (42), we obtain that
\bq
A=\gpavgs^{1/2}=2^{1/4}\mu_t^{-1/4}\left(\mf^{1/2}+\frac{\tan\a}{\mf^{1/2}}\right)^{1/2}.
\eq
which is consistent with the analytical result we obtained in (27).

\subsection{Optimal Curvature Analysis}
For a more general body shape, to satisfy the $y$-force balance and torque balance, we instead prescribe the curvature $\k(s,t)$
and obtain $x(s,t)$, $y(s,t)$, and $\theta(s,t)$ from equations (1)-(3).
We now correct the above analysis to satisfy all three equations. We again assume that the deflection from $x$-axis
is small and we decompose $y(s,t)$ as:
\bqa
y(s,t)&=& y_0(t)+\int_0^s\sin\th(s',t)ds'\\
 &=&y_0(t)+\int_0^s\th(s',t)ds'+O(y^3)\\
 &=&y_0(t)+\int_0^s\th_0(t)+\int_0^{s'}\k(s'',t)ds''ds'+O(y^3)\\
 &=&y_0(t)+s\th_0(t)+\int_0^s\int_0^{s'}\k(s'',t)ds''ds'+O(y^3)\\
 &\equiv &Y(t)+sR(t)+k(s,t)+O(y^3).
\eqa
Prescribing the curvature is the equivalent to prescribing $k(s,t)$. We set
\bq
k(s,t)=g(s+U_wt)
\eq
so $y$ is similar to the form given before, with two additional terms: vertical translation $Y(t)$ and rotation $R(t)$.
We determine $Y$ and $R$ by expanding the $y$-force (12) and torque balance (13) equations to leading order in $|y|$ and obtain:
\bqa
\int_0^1\left(-\frac{U_w}{U}+1+\frac{1}{2}\gpavgs\right)g'-\frac{Y'}{U}-\frac{R's}{U}+R\, ds =0,\\
\int_0^1s\left[\left(-\frac{U_w}{U}+1+\frac{1}{2}\gpavgs\right)g'-\frac{Y'}{U}-\frac{R's}{U}+R\right]\,ds =0.
\eqa
We solve (52) and (53) for $Y$ and $R$:
\bqa
\frac{Y'}{U}-R&=&4\Gamma-6\Lambda,\\
\frac{R'}{U}&=&12\Lambda-6\Gamma.\\
\Lambda&\equiv&\left(-\frac{U_w}{U}+1+\frac{1}{2}\gpavgs\right)\sg,\\
\Gamma&\equiv&\left(-\frac{U_w}{U}+1+\frac{1}{2}\gpavgs\right)\gp.
\eqa
We then express the $x$-force balance equation in terms of $Y$ and $R$ and obtain:
\bq
\int_0^1 -\cos\a\left[\mf+\mu_t \left(-\frac{U_w}{U}+1+\frac{1}{2}\gpavgs\right)g'^2+\mu_t\left(-\frac{Y'}{U}-\frac{R's}{U}+R\right)g'\right]-\sin\a \,ds=0.
\eq
Substituting (54) and (55) into (58) we solve for $U$ in terms of $g$:
\bq
\frac{U_w}{U}=1+\frac{1}{2}\gpavgs+\frac{\mf+\tan\a}{\displaystyle\mu_t\left(\gpavgs-\gp^2-3(\gp-2\sg)^2\right)}.
\eq
The cost of locomotion $\eta$ then becomes
\bq
\eta=1/\int_0^1\frac{1}{\displaystyle \sin\a+\cos\a\left(\mf\left(1+\frac{1}{2}\gpavgs\right)+\frac{(\mf+\tan\a)^2}{\displaystyle\mu_t\left(\gpavgs-\gp^2-3(\gp-2\sg)^2\right)}\right)}dt.
\eq

Following \cite{Alben}, we expand $g'$ in a basis of Legendre
polynomials $L_k$ for any fixed time $t$. The Legendre polynomials are
orthonormal functions with unit weight on $[0,1]$. They satisfy the relations
\bq
\int_0^1L_iL_jds=\delta_{ij};\ L_0 = 1,\ L_1=\sqrt{12}(s-1/2),\cdots
\eq
We write $g'$ as:
\bq
g'(s+U_wt)=\sum\limits_{k=0}^{\infty}c_k(t)L_k(s),
\eq
and we have
\bqa
\gpavgs=\sum\limits_{k=0}^{\infty}c_k(t)^2,\\
\gp^2=c_0(t)^2,\\
3(\gp-2\sg)^2=c_1(t)^2.
\eqa
Inserting into (60) we obtain
\bq
\eta=1/\int_0^1\frac{1}{\displaystyle\cos\a\left(1+\frac{1}{2}\left(c_0(t)^2+c_1(t)^2+\sum\limits_{k=2}^\infty c_k(t)^2\right)+\frac{ (\mf+\tan\a)^2}{\mu_t\sum\limits_{k=2}^\infty c_k(t)^2}\right)+\sin\a }dt.
\eq
Therefore, $\eta$ is minimized for
\bq
c_0(t)=0;\ c_1(t)=0;\ \sum\limits_{k=2}^\infty c_k(t)^2 = \sqrt{2}\mu_t^{-1/2}\left(\mf^{1/2}+\frac{\tan\a}{\mf^{1/2}}\right)
\eq
If a periodic function $g(s+U_wt)$ satisfies (67) for all $t$, this is the curvature function which
minimizes the cost of locomotion. The corresponding cost of locomotion is:
\bq
\eta\longrightarrow (\mf\cos\a+\sin\a)\left(1+\left(\frac{2\mf}{\mt}\right)^{1/2}\right).
\eq

Finally, we define the amplitude of the snake motion as
\bq
A\equiv\left(\frac{1}{U_w}\int_0^{U_w}g'(x)^2dx\right)^{1/2}.
\eq

Then we obtain that:
\bqa
\gpavgs&=&A^2+O(U_w),\\
\gp & =& O(U_w),\\
\sg &=& O(U_w).
\eqa

Therefore, in the limit of $U_w\rightarrow 0$, (67) holds with
the optimal
\bq
A=2^{1/4}\mu_t^{-1/4}\left(\mf^{1/2}+\frac{\tan\a}{\mf^{1/2}}\right)^{1/2}.
\eq The derivation of (70)-(72) can be found in an appendix of \cite{Alben}. We notice that when $\a=0$ and $\mf=1$, $\eta\rightarrow 1+\sqrt{2}\mu_t^{-1/2}$ and the optimal $A\rightarrow 2^{1/4}\mu_t^{-1/4}$, which are consistent with the optimal solutions of  snake motion on a level plane, derived in \cite{Alben}.

For the sinusoidal wave motion, we prescribed the curvature of the sinusoidal wave as
\bq
\k(s,t)=K\cos(n\pi s+2\pi t),
\eq
and according to (69), we obtain the amplitude $A$
\bq
A=\frac{K}{\sqrt{2}n\pi}.
\eq
Therefore, the magnitude of the sinusoidal wave $K$ scales as \bq
K=(\sqrt{2}n\pi)(2^{1/4}\mu_t^{-1/4})\left(\mf^{1/2}+\frac{\tan\a}{\mf^{1/2}}\right)^{1/2}.\eq
which is consistent with our numerical results.

We now discuss the case in which the snake's net displacement is not solely
in the $x$ direction, up the incline,
but instead has a nonzero component in the $y$ direction, across the incline.
The above analysis and \cite{Alben} show that in the limit of large
$\mt$, the minimum cost of locomotion is achieved when the curvature is any traveling
wave function, in the limit of vanishing wavelength, and with amplitude tending
to zero like $\mt^{-1/4}$. For all such optimal motions, the snake
moves along a straight-line path.
Let the distance travelled by the center of mass over
one period be $d$, with $x$- and $y$-displacements $d_x$ and $d_y$,
so $d = \sqrt{d_x^2 + d_y^2}$.
We may redefine $\eta$ as $W/d_x$ now, so only the distance travelled up the incline is considered
useful. We also set $\eta = +\infty$ if $d_x < 0$, to avoid the trivial case in
which the snake travels down the incline, which we discussed in Section II.
Our previous results continue to hold with this definition of $\eta$, because we set the
initial orientation of the snake so that its center of mass travels only in the $x$ direction and $d = d_x \geq 0$ in all cases. Now if $d_y \geq 0$,
 we first claim that the optimal
motions in the large-$\mt$ limit still follow straight-line paths. Any non-straight path
with the same beginning and end points would have a greater arc length, and thus
require more work done against forward friction for the same distance travelled by the snake's center of mass. For a straight-line path, the work against
transverse friction vanishes (corresponding to $\eta$ in
equation (43) in the limit of large $\mt$), so it could
not be decreased for the non-straight path. Work against gravity is the same for the
straight and non-straight paths since $d_x$ is the same. Therefore, the straight-line path is the optimal path to minimize $\eta$. Now we show that the
$\eta$-minimizing path is a straight-line path with $d_y = 0$. $\eta$ is now
a generalized version of equation (43):
\bq
\eta\rightarrow\mu_f\cos\alpha\frac{d}{dx}+\sin \alpha,\quad \mbox{as}\quad  \mt\rightarrow\infty,
\eq
and it is minimized when $d_y = 0$ and $d = d_x$. In short, nonzero $d_y$ increases the work
against forward tangential friction without any compensating increase in $d_x$, so $\eta$ increases.

\section{CONCLUSION}
We have studied the optimization of the snake motions on an inclined plane. We used a two-dimensional model and determined the effects of the
parameters---the transverse and tangential coefficients of friction and the incline angle---on the optimal shape for
triangular wave motion and sinusoidal wave motion. When the transverse friction coefficient is much larger than the tangential friction coefficient,
we showed that for a given incline angle $\a$ and tangential friction coefficient $\mf$, the cost of locomotion tends to
$(\mf\cos\a+\sin\a)$ with the optimal amplitude scaling as $\mt^{-1/4}$. Our analysis also showed a non-monotonic relationship between the cost of locomotion and the incline angle. The least
efficient optimal motion is achieved at a critical incline angle depending on the value of $\mf$. The optimal amplitude increases with the incline angle to decrease slipping. For given $\mt$ and $\a$, the motion becomes less efficient as $\mf$ increases due to the extra work against tangential friction. However,
when $\mf$ is small, we find that motion with a larger amplitude is more efficient, while when $\mf$ is large, a motion with smaller amplitude is optimal. We gave an asymptotic analysis allowing a more general class of motions and our asymptotic results showed the same scaling laws for optimal amplitude and cost of locomotion.

An extension of this work is to discuss three-dimensional motions of snakes and include wider parameter spaces with small or moderate transverse friction coefficients. Another interesting direction is to consider motions in the presence of walls \cite{Marvi,Maneewarn,Transeth}.

\end{document}